\documentclass[aps,prl,amsmath,amssymb,reprint,superscriptaddress,nofootinbib]{revtex4-1}

\usepackage{graphicx}
\usepackage{dcolumn}
\AtBeginDocument{\usepackage{booktabs}}
\usepackage{units}
\usepackage{hyperref}
\usepackage{braket}

\usepackage{color}
\definecolor{red}{RGB}{255,0,0}

\usepackage{comment}

\begin{document}
	

\newcommand{\Be}{\textsuperscript{9}Be\textsuperscript{+}}
\newcommand{\Mg}{\textsuperscript{25}Mg\textsuperscript{+}}
\newcommand{\Mgg}{\textsuperscript{24/25}Mg\textsuperscript{+}}
\newcommand{\Mgt}{\textsuperscript{24}Mg\textsuperscript{+}}
\title{Electromagnetically-Induced-Transparency Cooling with a Tripod Structure in a Hyperfine Trapped Ion with Mixed-Species Crystals}

\author{Jenny J. Wu}
\affiliation{National Institute of Standards and Technology, 325 Broadway, Boulder, CO 80305, USA}
\affiliation{Department of Physics, University of Colorado, Boulder, CO 80309, USA}

\author{Pan-Yu Hou}
\affiliation{National Institute of Standards and Technology, 325 Broadway, Boulder, CO 80305, USA}
\affiliation{Department of Physics, University of Colorado, Boulder, CO 80309, USA}
\altaffiliation[]{Current address: Institute for Interdisciplinary Information Sciences, Tsinghua University, Beijing, China}

\author{Stephen D. Erickson}
\affiliation{National Institute of Standards and Technology, 325 Broadway, Boulder, CO 80305, USA}
\affiliation{Department of Physics, University of Colorado, Boulder, CO 80309, USA}
\altaffiliation[]{Current address: Quantinuum, 303 South Technology Court, Broomfield, Colorado 80021, USA}

\author{Adam D. Brandt}
\affiliation{National Institute of Standards and Technology, 325 Broadway, Boulder, CO 80305, USA}
\affiliation{Department of Physics, University of Colorado, Boulder, CO 80309, USA}
\altaffiliation[]{Current address: Vapor Cell Technologies, Boulder, Colorado, USA}

\author{Yong Wan}
\affiliation{National Institute of Standards and Technology, 325 Broadway, Boulder, CO 80305, USA}
\affiliation{Department of Physics, University of Colorado, Boulder, CO 80309, USA}
\altaffiliation[]{Current address: University of Science and Technology of China, Hefei, China}

\author{Giorgio Zarantonello}
\affiliation{National Institute of Standards and Technology, 325 Broadway, Boulder, CO 80305, USA}
\affiliation{Department of Physics, University of Colorado, Boulder, CO 80309, USA}
\altaffiliation[]{Current address: Qudora Technologies GmbH, Braunschweig, Germany}

\author{Daniel C. Cole}
\affiliation{National Institute of Standards and Technology, 325 Broadway, Boulder, CO 80305, USA}
\altaffiliation[]{Current address: Infleqtion, 3030 Sterling Circle, Boulder, CO 80301, USA}

\author{Andrew C. Wilson}
\affiliation{National Institute of Standards and Technology, 325 Broadway, Boulder, CO 80305, USA}

\author{Daniel H. Slichter}
\affiliation{National Institute of Standards and Technology, 325 Broadway, Boulder, CO 80305, USA}

\author{Dietrich Leibfried}
\affiliation{National Institute of Standards and Technology, 325 Broadway, Boulder, CO 80305, USA}

\begin{abstract}
Cooling of atomic motion is a crucial tool for many branches of atomic physics, ranging from fundamental physics explorations to quantum information and sensing. For trapped ions, electromagnetically-induced-transparency (EIT) cooling has received attention for the relative speed, low laser power requirements, and broad cooling bandwidth of the technique. However, in applications where the ion used for cooling has hyperfine structure to enable long coherence times, it is difficult to find a closed three-level system in which to perform standard EIT cooling. Here, we demonstrate successful EIT cooling on \Mg\, by the addition of an extra laser frequency; this method can be applied to any ion with non-zero nuclear spin. Furthermore, we demonstrate simultaneous EIT cooling of all axial modes in mixed-species crystals \Be-\Mg\, and \Be-\Mg-\Be\, through the \Mg\, ion.
\end{abstract}
\maketitle

\section{Introduction}
Control over the motion of atoms and molecules is an important prerequisite for many applications including precision spectroscopy~\cite{RevModPhysAtomicClocks}, quantum control and sensing~\cite{RevModPhysQuantumSensing}, and quantum information processing~\cite{QuantumComputingReview, QuantumComputingIonReview, Bruzewicz2019}.
Uncooled atomic motion introduces systematic effects in precision measurements that must be carefully characterized~\cite{RevModPhysAtomicClocks}, and it can also limit the fidelity of quantum gates~\cite{QuantumComputingIonReview, Bruzewicz2019}.
While early experiments involving atomic ions utilized Doppler laser cooling~\cite{WinelandLaserCooling, DehmeltDopplerCooling}, modern experiments often benefit from 
reaching\textemdash or require\textemdash sub-Doppler thermal occupation of some or all ion normal modes of motion~\cite{RevModPhysAtomicClocks, RevModPhysQuantumSensing, QuantumComputingIonReview}. 

Commonly used sub-Doppler cooling methods for ion trapping include resolved-sideband cooling~\cite{Diedrich1989, Monroe1995, LaserCoolingEschner}, Sisyphus cooling~\cite{CiracPGCTheory1993, JavanainenPGCTheory1993, SisyphusWineland, Birkl1994, Sisyphus2017}, and electromagnetically-induced-transparency (EIT) cooling~\cite{MorigiEIT2000, MorigiEIT2003, EITCaRoos, YihengMgEIT, Lechner2016, EITPenning, EITYbMonroe, EITYbKim}. 
For experiments requiring approximate ground states of motion to be prepared, Sisyphus cooling is not sufficient as it can only achieve a mean thermal occupation of order unity. Resolved sideband cooling often takes up a substantial portion of the experimental duty cycle~\cite{Bowler2013, GateTeleportation, StephenNonresonant, QuantinuumRacetrack}, so faster cooling methods are desirable. Faster cooling reduces the total experiment duration and the steady-state motional occupation in the presence of motional heating, and increases the amount of data that can be taken between recalibrations when experimental parameters have slow drifts. EIT cooling is an attractive candidate, as it is relatively fast, can cool multiple modes simultaneously, uses lower power than Raman sideband cooling, and can achieve a lower mode occupation than Sisyphus cooling \cite{MorigiEIT2000, MorigiEIT2003}.

In trapped ions, EIT cooling has been demonstrated using several species including \textsuperscript{40}Ca\textsuperscript{+}~\cite{EITCaRoos}, \Mgt~\cite{YihengMgEIT}, \Be~at a magnetic field of 4.46 T where the electron spin is effectively decoupled from the nuclear spin~\cite{EITPenning}, and more recently \textsuperscript{171}Yb\textsuperscript{+}~\cite{EITYbMonroe, EITYbKim}. Here we extend the technique to \Mg, an ion that has nuclear spin $I=5/2$ and thus a large hyperfine ground state manifold. The number of hyperfine states makes \Mg~susceptible to population leakage from laser scatter during cooling, which makes it harder to reach a motionally cold steady state. However, the hyperfine structure also allows for the possibility of transitions between ground states whose frequencies are first-order magnetic-field-insensitive at finite magnetic fields, affording long coherence times~\cite{Bollinger1991, LongLivedQubit, Harty2014, Sepiol2019}. In experiments where the ion species used for cooling is also required to participate in coherent operations such as indirect readout or state preparation via quantum logic spectroscopy~\cite{SchmidtQLS,StephenNonresonant} or entanglement distribution~\cite{GateTeleportation}, such field-insensitive transitions can be beneficial.

Here, we address the issues arising from the hyperfine structure, which makes it hard to isolate three states in a $\Lambda$-system as traditionally used in EIT cooling, by using an additional laser field. It should be possible to use this approach in other ion species with nonzero nuclear spin.
We also demonstrate sympathetic cooling of various mixed-species \Be and \Mg\,ion crystals by EIT cooling of \Mg, including indirect cooling of a mode in a \Be-\Mg-\Be\,crystal in which \Mg\,does not participate by combining EIT cooling with parametric mode coupling~\cite{Modecouplingforcooling}. 

\section{Cooling Using EIT}
EIT cooling in an ideal three-level system is described in Refs.~\cite{MorigiEIT2000, MorigiEIT2003, RevModPhysDidi}. The EIT effect in its simplest form results from modifying the absorption profile of a three-level system through quantum-mechanical interference. This is often done by dressing the system with two laser beams. If the laser parameters satisfy the EIT-cooling relation (stated below) and atomic motion is considered, the resulting laser absorption profile is greatly enhanced at the frequency corresponding to the loss of motional quanta, while motion-preserving (carrier) scatter is suppressed. The resulting asymmetry between the motion-adding and motion-subtracting scatter (favoring the latter) yields a low steady-state thermal occupation. 

\begin{figure}[!h]
    \centerline{\includegraphics[width=0.45\textwidth]{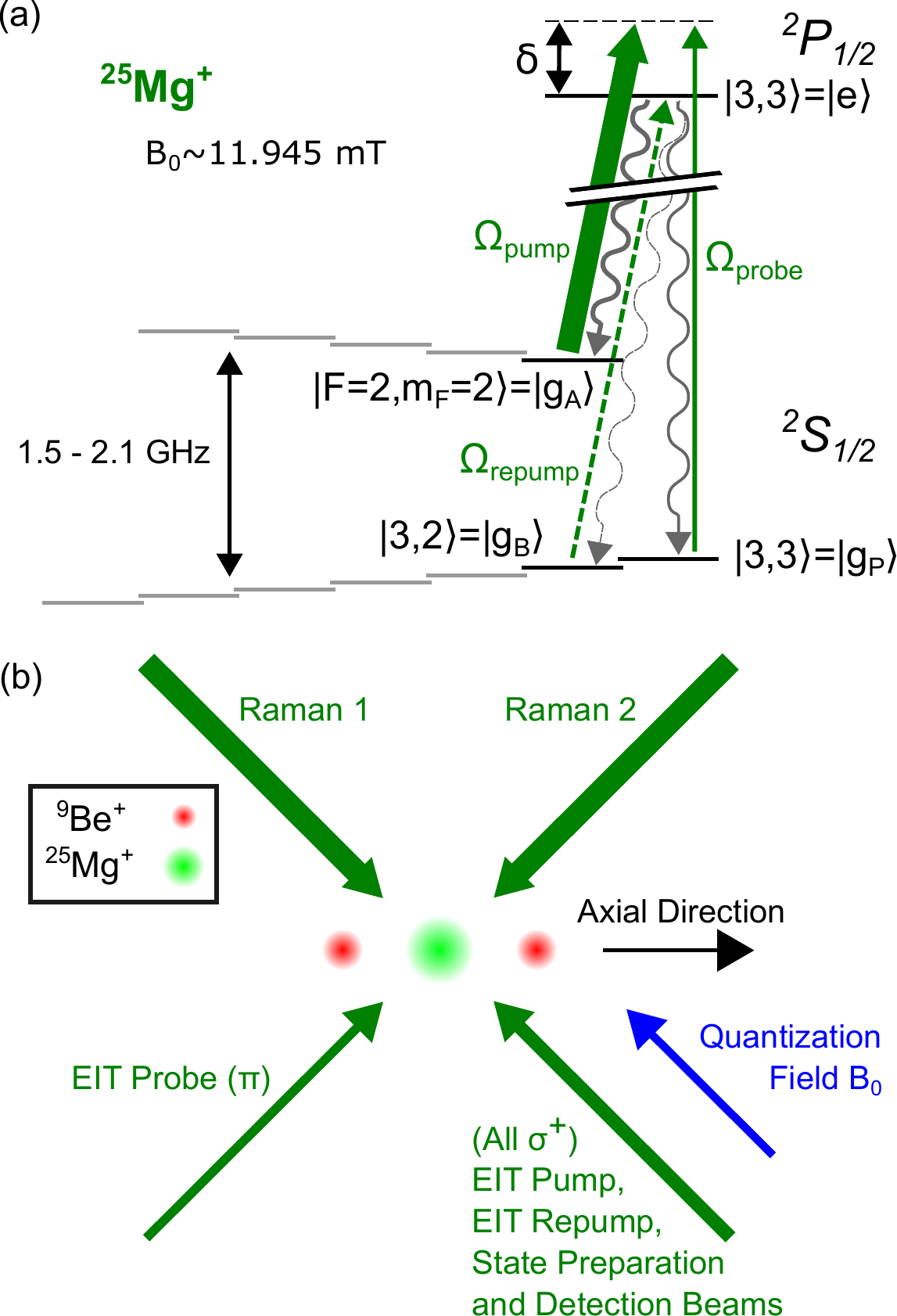}}
    \caption{Energy level diagram for \Mg, laser frequencies, polarizations, and beam orientations. (a) shows EIT cooling excitation and decay channels for \Mg. The $\Lambda$-structure of EIT contains the states labeled $\ket{g_A}$, $\ket{g_P}$, and $\ket{e}$, while the presence of extra state $\ket{g_B}$ requires a repump beam to $\ket{e}$. Green arrows represent laser fields, and wavy grey lines represent decay via spontaneous emission; the solid lines and decays are part of the EIT $\Lambda$ system while the dashed lines indicate the repump laser from and decay to the extra state. The relative strengths of the laser fields and branching ratio of the decays are indicated qualitatively by the thicknesses of the arrows. The two laser beams involved in EIT are blue-detuned from resonance with detuning $\delta$ (see main text). (b) shows the geometry of relevant \Mg\,beams used; all lasers are at $\pm 45 ^\circ$ to the axial direction along which ion chains are oriented; a \Be - \Mg - \Be~crystal is sketched as an example (\Be\, in red, \Mg\, in green). The EIT probe and pump beams (as well as the Raman 1 and 2 beams) have a wavevector difference that is approximately aligned to the axial direction, making this configuration well-suited for cooling and characterizing the axial motional modes.}
    \label{fig: EITtheory}
\end{figure}

Typically, a three-level system with two ground states and an excited state in a ``$\Lambda$" structure is considered. The dressing lasers are usually a strong ``pump'' laser beam and a weak ``probe'' laser beam that drive transitions from each of the two ground states.  
In order to perform cooling, these lasers are blue-detuned from resonance with the transitions from their respective ground state to the upper state, and the wavevector difference of the two beams must have a nonzero projection along the direction of the mode(s) that are to be cooled (see Fig.~\ref{fig: EITtheory}(b)). 

The minimal motional occupation is reached when $\nu = (\Omega_{\rm pump}^2 + \Omega_{\rm probe}^2)/{4\delta}$, where $\nu$ is the angular frequency of the motional mode of interest, $\Omega_{\rm pump}$ and $\Omega_{\rm probe}$ express the intensities of the pump and probe beams in terms of equivalent resonant Rabi frequencies (see Appendix B), and $\delta$ is the common detuning of the pump and probe laser frequencies from their respective atomic resonances. The theoretical minimum thermal occupation achievable in an ideal three-level $\Lambda$-system is given by $(\Gamma/(4\delta))^2$, where $\Gamma$ is the linewidth of the excited state.

An isolated subsystem with a $\Lambda$ configuration or equivalent for EIT is hard to realize in \Mg~(and similar ions) with its many hyperfine states. Instead, it is necessary to consider at least three ground states and one excited state that form a ``tripod" structure, as shown in Fig.~\ref{fig: EITtheory}(a). To perform EIT cooling in a tripod subsystem, it is advantageous to prevent population from accumulating in the ``extra'' ground state ($\ket{g_B}$ in Fig.~\ref{fig: EITtheory}(a)), for example by adding a third, weak beam to repump population. The additional repump beam and the strong pump beam both have $\sigma^+$ polarization to keep the population in the tripod subsystem. This method was demonstrated successfully with $^{171}$Yb$^+$~ ions~\cite{EITYbMonroe, EITYbKim}. The cooling rate is slower and final thermal occupation is higher than for the ideal $\Lambda$-system case. For the remainder of this paper, we refer to this tripod-system method simply as ``EIT cooling", even though it uses an additional repump beam. 
A second scheme for EIT cooling in a tripod level structure involves adding a strong third beam with similar detuning to the pump and probe beams, thereby performing ``double EIT" cooling as proposed in Ref.~\cite{DoubleEIT}.
While we were able to achieve cooling using this method as well, it is much less effective than expected due to experimental issues specific to our apparatus, most notably uncompensatable axial micromotion (which also has a substantial effect on the EIT scheme with weak repump). More detail on double EIT is available in Appendix A.

\section{Experimental Implementation}
Experiments are performed in a room-temperature rf Paul trap with linearly segmented DC electrodes~\cite{BlakestadTrap}. The apparatus is used to trap both \Be~and \Mg~ions, and the direction of weakest confinement is labeled the ``axial" direction. The motional mode frequencies for the axial and two radial modes are $2\pi\times\{1.18, 4.4, 4.6\}$ MHz, respectively, for a single \Mg~ion and $2\pi\times\{1.97, 12.9, 13.0\}$ MHz for a single \Be~ion.
Due to the trap design, the ions experience an axial rf electric field at $\omega_{rf} = 2\pi \times 82.5$ MHz, resulting in unavoidable axial micromotion as detailed below and in Ref.~\cite{BlakestadThesis}.

We use a magnetic quantization field of 11.945 mT, oriented as shown by the blue arrow in Fig.~\ref{fig: EITtheory} (b), which renders the \textsuperscript{2}S\textsubscript{1/2}$\ket{F=2, m_F=0}\leftrightarrow\ket{F=1,m_F=1}$ transition frequency in \Be~first-order insensitive to magnetic field fluctuations. \Mg~is co-trapped with \Be~and is used for sympathetic cooling and the characterization of motional states in this work.
At this field, the Zeeman splittings between neighboring $m_F$ states within each $F$ manifold of the \textsuperscript{2}S\textsubscript{1/2} ground level range from 48 MHz to 66 MHz, while the states within the $F=2$ and $F=3$ manifolds are separated by between 1.520 and 2.075 GHz.
The internal state of \Mg\,is prepared using optical pumping to state \textsuperscript{2}S\textsubscript{1/2}$\ket{3,3}$ ($\ket{g_P}$ in Fig.~\ref{fig: EITtheory} (a)), followed by microwave pulses as necessary to move population to a different $\ket{F,m_F}$ state. Sideband interactions that couple the internal state of \Mg\,and the motion of the ions are driven by Raman laser beams detuned from resonance with the nearest atomic transition by approximately 
$2\pi \times 200$ GHz; these can be used for resolved sideband cooling and for characterization of the states of motion~\cite{ExperimentalIssues, RevModPhysDidi}. The internal states of the ions are distinguished by transferring the populations of two relevant states to either the ``bright" (maximally fluorescing) state or a ``dark" (minimally fluorescing) state using microwave pulses, and then applying fluorescence detection~\cite{Dehmelt1982, Wineland1987, CalciumFluorescenceNagerl1998, CoolingcollectivemotionKing1998, ExperimentalIssues, ExperimentalPrimer, RevModPhysDidi}. 

The EIT beams are all derived from a single laser source addressing the \textsuperscript{2}S\textsubscript{1/2} $\leftrightarrow$ \textsuperscript{2}P\textsubscript{1/2} transition that is also used for optical pumping. The source is split into three beams that we call ``pump'' ($\sigma^+$-polarized, driving the \textsuperscript{2}S\textsubscript{1/2}$\ket{2,2}$ $\rightarrow$ \textsuperscript{2}P\textsubscript{1/2} $\ket{3,3}$ transition), ``repump'' ($\sigma^+$, \textsuperscript{2}S\textsubscript{1/2} $\ket{3,2}$ $\rightarrow$ \textsuperscript{2}P\textsubscript{1/2} $\ket{3,3}$), and ``probe'' ($\pi$, \textsuperscript{2}S\textsubscript{1/2} $\ket{3,3}$ $\rightarrow$ \textsuperscript{2}P\textsubscript{1/2} $\ket{3,3}$), as shown in Fig.~\ref{fig: EITtheory}(a). Each beam is controlled by separate acousto-optic modulators, enabling independent control of beam powers and frequencies. The pump and probe beams are used to form the EIT resonance with a common (blue) detuning of $\delta\approx 2 \pi \times 180$ MHz, while the repump beam is held on resonance with the \textsuperscript{2}S\textsubscript{1/2}$\ket{3, 2}$ $\leftrightarrow$ \textsuperscript{2}P\textsubscript{1/2} $\ket{3, 3}$ transition. 

Optimal cooling parameters were found by cooling for a fixed period of time\textemdash typically 5 ms to ensure the steady state is reached\textemdash and iteratively varying the power of each beam. More details on the calibration are available in Appendix B.

\section{EIT Cooling of a Single Magnesium Ion}
\begin{figure}[!h]
    \centerline{\includegraphics[width=0.48\textwidth]{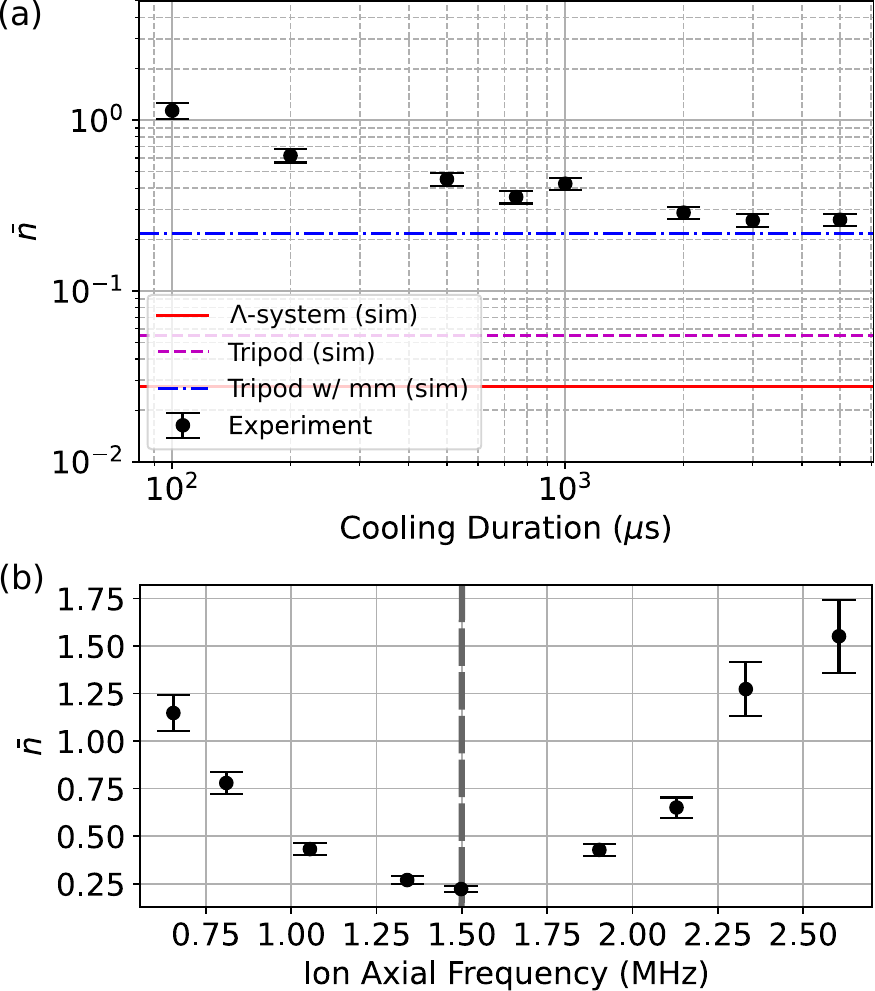}}
    \caption{EIT cooling of a single \Mg~ion. (a) Cooling dynamics of the axial mode of a \Mg~ion to steady-state average occupation $\bar{n} =0.26\pm0.02$ over approximately 1.5 ms. Simulations of the steady-state $\bar{n}$ for a $\Lambda$-EIT system (red solid line), a tripod-EIT system (magenta dashed line), and a tripod-EIT system with micromotion (blue dash-dot line) are shown compared with the experimentally determined $\bar{n}$ (black dots). (b) Motional mode occupation after 3 ms of EIT cooling for varying axial mode frequencies of a single \Mg~using EIT cooling parameters calibrated for a mode frequency of 1.5 MHz, indicated by the vertical dashed line. The cooling is observed to be effective over a bandwidth of $\approx1.5$ MHz around this center frequency, with higher final $\bar{n}$ for mode frequencies further from 1.5 MHz.}
    \label{fig: SingleMgEIT}
\end{figure}

First, we demonstrate EIT cooling on the $\nu= 2\pi \times$1.18 MHz axial mode of a single \Mg\, ion. The ion is first Doppler-cooled to an average occupation number $\bar{n}\approx 10$ and then EIT cooling is performed for varying durations. The beam intensities used for this experiment, expressed as equivalent resonant Rabi frequencies, are estimated (see Appendix B for details) to be $\Omega_{\rm pump}/2\pi = 29.38 \pm 0.21$ MHz, $\Omega_{\rm probe}/2\pi= 5.41\pm0.09$ MHz, and $\Omega_{\rm repump}/2\pi=0.87\pm 0.06$ MHz. Data were taken only when $\bar{n}$ was close to one to avoid excessive uncertainty when estimating higher $\bar{n}$ values using sideband thermometry \cite{RevModPhysDidi}. The resulting average occupation $\bar{n}$ is shown in Fig.~\ref{fig: SingleMgEIT}(a) together with results from three different master-equation simulations of the system with the Quantum Toolbox in Python (QuTiP) \cite{johansson2012qutip}. 

The simulations are performed based on different models: a standard $\Lambda$-EIT system; an ideal tripod-EIT system; and a tripod-EIT system including the effects of micromotion, which gives rise to sidebands on the driving light field as described below. All simulations incorporate the same experimentally measured motional heating rate of 320 quanta/s and the recoil from scattered photons. The two models of tripod systems use the estimated $\Omega$ values, while the $\Lambda$-EIT model uses the experimentally estimated $\Omega_{\rm probe}$ and a calculated ideal $\Omega_{\rm pump}$. The model that includes micromotion also incorporates an extra detuning beyond $\delta = 2\pi \times 180$ MHz of $2\pi \times 0.75$ MHz of the probe beam to account for Stark shifts due to off-resonant micromotion sidebands. In our experiment, it was found that optimal cooling required such an extra detuning of the probe beam. 

The experimentally determined steady-state occupation number of the ion greatly exceeds that found in the $\Lambda$-EIT model. While the addition of a third level in the ideal tripod-EIT simulation reduces the discrepancy, the results of the model including micromotion are the closest to experimental results, suggesting that micromotion has a major impact on the observed steady state occupation.

For an ion experiencing micromotion, laser light in the ion frame is phase-modulated at the trap rf frequency, resulting in sidebands on all incident beams. These sidebands have frequencies $\omega_L + n\,\omega_{rf}$ where $\omega_L$ is the carrier laser frequency, $\omega_{rf}$ is the trap rf frequency, and $n$ is the integer sideband order. The sidebands have relative field amplitudes scaled by the Bessel function $J_{|n|}(\beta)$, where $\beta$ is the micromotion modulation index \cite{MicromotionMinimization}.

For the EIT pump and probe beams, some of the sidebands with $n<0$, despite being weaker than the carrier ($n=0$), are closer to resonance with the ion's optical transitions, making their effects non-negligible during EIT cooling. Recoil from photon scattering on these sidebands can cause heating and slow down progress towards the system's steady state, while the Stark shifts they cause can complicate optimization of EIT parameters (see Appendix B). For our case, $\beta=0.827$ is measured independently using Raman micromotion sideband oscillations. The $n=-1$ and $n=-2$ micromotion sidebands of both the pump and probe beams have detunings from resonance of $2\pi \times 97.5$ MHz and $2\pi \times 14.9$ MHz, and powers relative to the carrier of $\approx$ 20\% and $\approx$ 1\%, respectively.

Using experimental parameters that were measured independently, the simulated steady-state axial mode $\bar{n}$ for the $\Lambda$, tripod, and tripod-with-micromotion systems were found to be $\bar{n}_{\rm \Lambda}=0.028$, $\bar{n}_{\rm tripod}=0.055$, and $\bar{n}_{\rm MM}=0.216$ respectively. The difference between the experimentally measured occupation of $\bar{n}_{exp}=0.26\pm0.02$ and $\bar{n}_{\rm MM}$ is attributed to additional experimental imperfections, such as beam power fluctuations resulting from beam pointing drifts and AOM temperature instability.

Although the presence of uncompensatable micromotion in our trap affects the final occupation, cooling well below $\bar{n}=1$ is achieved within 500 $\mu$s. After this duration, we find $\bar{n}\lesssim 0.3$. An even lower occupation can be reached with a short period of sideband cooling after EIT cooling, potentially with a shorter overall duration than resolved sideband cooling starting at the Doppler limit.

The larger bandwidth of EIT cooling compared to resolved sideband cooling is one of the most compelling features of this technique. Some previous demonstrations focused on the use of EIT cooling to simultaneously cool many modes spanning a frequency range up to multiple MHz \cite{Lechner2016, EITPenning, EITYbKim}. Here, we take a different approach and vary the axial mode frequency of a single ion to probe the cooling bandwidth. The cooling is first optimized for a mode frequency of $2\pi \times 1.5$ MHz. For this setting, the optimized parameters are found to be $\Omega_{\rm pump}/2 \pi= 27.19\pm0.21$ MHz, $\Omega_{\rm probe}/2 \pi= 4.45\pm0.06$ MHz, and $\Omega_{\rm repump}/2 \pi = 0.70\pm 0.15$ MHz. Then the confining potential is varied in order to change the ion axial mode frequency while the EIT cooling parameters are held constant; the ion is cooled for 3 ms at each new frequency to approach steady state. The results are shown in Fig.~\ref{fig: SingleMgEIT}(b). The cooling is optimal near the frequency at which the EIT parameters were calibrated, but cooling is observed to be effective for a bandwidth of $\approx$ $2\pi \times 1.5$ MHz around the center frequency. 

\section{Sympathetic Multimode EIT Cooling of Mixed-Species Ion Chains}
\begin{figure}
    \centerline{\includegraphics[width=0.45\textwidth]{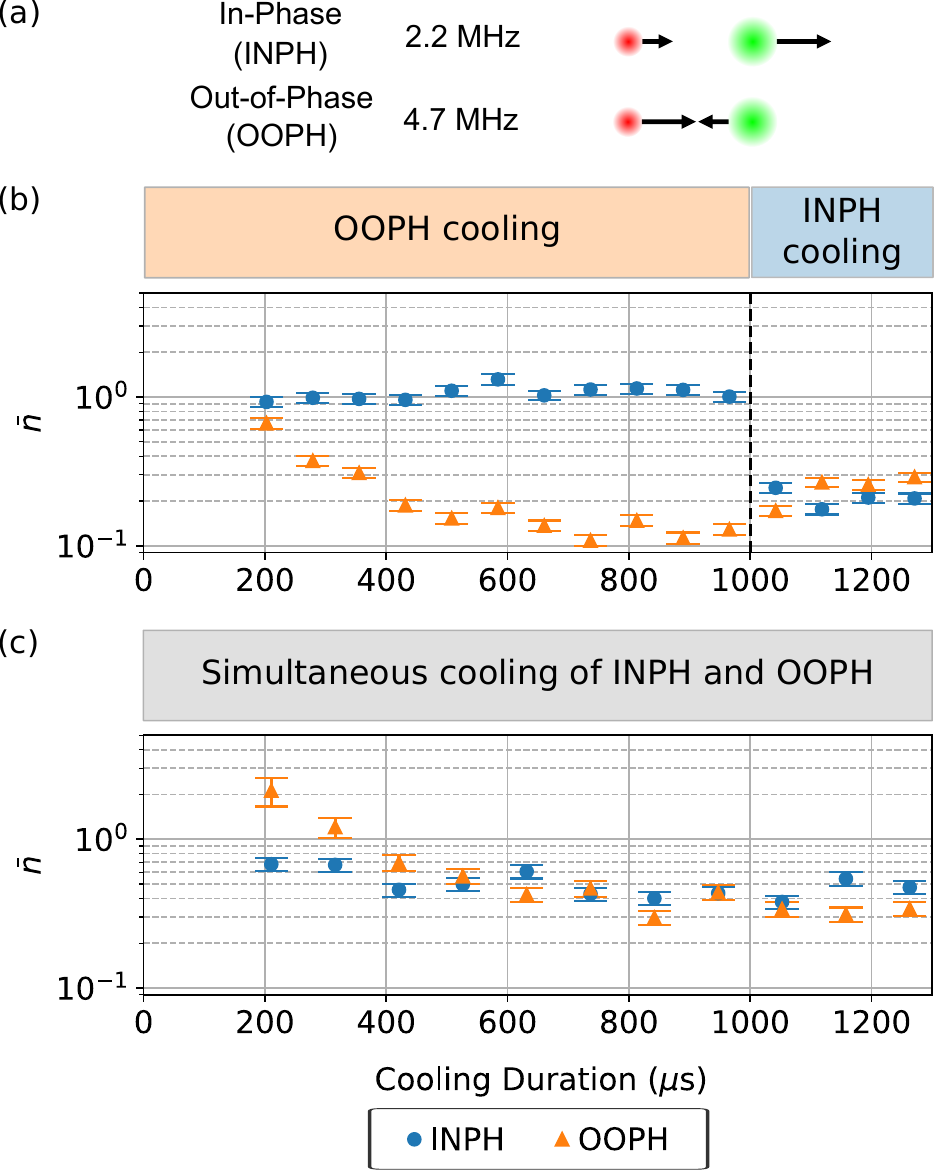}}
    \caption{EIT cooling of \Be-\Mg\, ion crystal. (a) shows the frequencies of the cooled modes as well as the participation of each ion (\Be\, in red, \Mg\, in green) in the respective mode in black arrows. (b) shows interleaved cooling, with 1000 $\mu$s of cooling optimized for the OOPH mode  followed by 300 $\mu$s of cooling optimized for the INPH mode. (c) shows simultaneous cooling with a single EIT parameter set. Cooling sequences for (b) and (c) are described in colored blocks above each plot.}
    \label{fig: MultimodeEITBeMg}
\end{figure}

\begin{figure}
    \centerline{\includegraphics[width=0.45\textwidth]{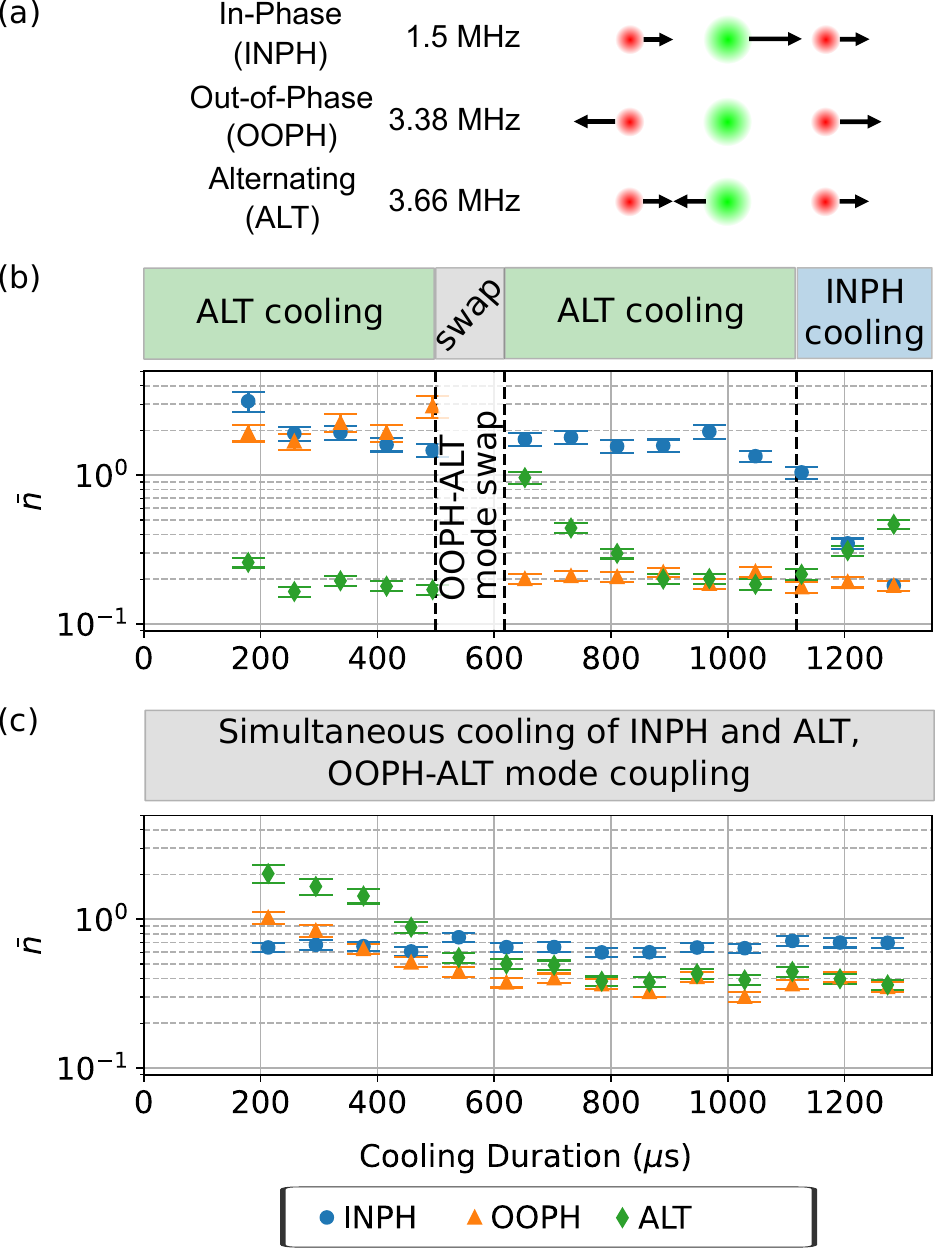}}
    \caption{EIT cooling of \Be-\Mg-\Be\, ion crystal. (a) shows the frequencies of the cooled modes as well as the participation of each ion (\Be\, in red, \Mg\, in green) in the respective mode in black arrows. (b) shows interleaved cooling, with 500 $\mu$s of ALT-mode-optimized cooling, a coherent swap between the OOPH and ALT modes taking 117 $\mu$s, 500 $\mu$s of ALT-mode-optimized cooling again, and finally 183 $\mu$s of INPH-mode-optimized cooling. (c) shows simultaneous cooling of all three axial modes. The coupling between the OOPH and ALT modes is turned on simultaneously with an EIT pulse that is optimized to cool all three modes. Cooling sequences for (b) and (c) are described in colored blocks above each plot.}
    \label{fig: MultimodeEITBeMgBe}
\end{figure}

Next, we demonstrate cooling of multiple axial modes of two different mixed-species ion crystals through \Mg. The crystals we consider here are \Be-\Mg\, with in-phase (INPH) and out-of-phase (OOPH) axial modes (mode frequencies are $2\pi \times 1.36$ MHz and  $2\pi \times 2.83$ MHz, respectively); and \Be-\Mg-\Be with in-phase (INPH), out-of-phase (OOPH), and alternating (ALT) axial modes (mode frequencies of $2\pi \times 1.50$ MHz, $2\pi \times 3.38$ MHz, and  $2\pi \times 3.66$ MHz, respectively). The motion of each ion in each of these modes are sketched in Figs.~\ref{fig: MultimodeEITBeMg}(a) and \ref{fig: MultimodeEITBeMgBe}(a). 

For both crystals, we attempt two different methods of cooling which we will call ``interleaved cooling" and ``simultaneous cooling". Interleaved cooling involves optimizing separate EIT parameters for each mode and then pulsing the cooling beams with each of these parameter sets in an alternating fashion, while simultaneous cooling involves optimizing a single set of parameters and relying on the broad bandwidth of EIT cooling to simultaneously cool all modes, as previously demonstrated in \cite{EITCaRoos, EITYbMonroe, EITYbKim, EITPenning}. Beam intensities used in each experiment are given in Appendix B.

Usually, for a reasonable EIT cooling parameter set, each mode has an equilibrium phonon occupation that is lower than the mode's Doppler-cooled occupation, so that some cooling of several modes can be performed. However, when the interleaved method is used, cooling a second mode $b$ after the first mode $a$ will cause mode $a$ to heat up towards its equilibrium occupation given the second parameter set that is optimized for mode $b$, as seen in Figs.~\ref{fig: MultimodeEITBeMg}(b) and \ref{fig: MultimodeEITBeMgBe}(b) at the end of the cooling sequence. Thus, we typically stop the last EIT pulse before the mode occupations reach equilibrium to achieve a lower final occupation for all modes. 

As seen in Figs. \ref{fig: MultimodeEITBeMg} and \ref{fig: MultimodeEITBeMgBe}, the simultaneous method cools all modes more quickly, but cannot achieve the same minimum occupation for each mode compared to the interleaved method, since the cooling is not optimized for any of the modes. For \Be-\Mg, the interleaved method achieved $\bar{n}_{\rm INPH} = 0.21\pm0.02$ and $\bar{n}_{\rm OOPH} = 0.29\pm0.02$, while the simultaneous method achieved $\bar{n}_{\rm INPH} = 0.46\pm0.05$ and $\bar{n}_{\rm OOPH} = 0.39\pm0.04$, where the subscripts correspond to the mode whose occupation is measured. For \Be-\Mg-\Be, the interleaved method achieved $\bar{n}_{\rm INPH} = 0.35\pm0.03$, $\bar{n}_{\rm OOPH} = 0.19\pm0.02$, and $\bar{n}_{\rm ALT} = 0.31\pm0.03$ (in this case taking the points closest to 1200 $\mu$s as the ``final'' occupations), while the simultaneous method achieved $\bar{n}_{\rm INPH} = 0.70\pm0.05$, $\bar{n}_{\rm OOPH} = 0.35\pm0.03$, and $\bar{n}_{\rm ALT} =0.36\pm0.03$.

In general, when performing sympathetic cooling it is necessary to consider the coupling of the coolant ion(s) to the mode(s) to be cooled. For example, in the OOPH mode in \Be-\Mg-\Be, the two \Be\, ions move symmetrically in opposite directions while the \Mg\, ion does not participate (see Fig.~\ref{fig: MultimodeEITBeMgBe} (a)), meaning this mode cannot be directly cooled using \Mg. To perform EIT cooling of all axial modes of this crystal using \Mg, we modulate the trapping potential at the frequency difference between the OOPH and ALT modes to couple these modes and exchange their occupations~\cite{Gorman2014, ModeCoupling, Modecouplingforcooling}. To perform sideband thermometry on the OOPH mode, the OOPH motional state is swapped to the ALT mode and characterized using sideband pulses on the \Mg~ion. For interleaved cooling, the swap operation is considered to be part of the cooling sequence, taking 117 $\mu$s, as shown in Fig.~\ref{fig: MultimodeEITBeMgBe}(b). It can also be seen that the occupation of the OOPH mode barely changes (apart from the swap) during the cooling sequence; in addition to being a mode with a low heating rate, it is also nearly immune to heating from \Mg~recoil because of the lack of participation in the OOPH mode~\cite{ModeCoupling}. When performing simultaneous cooling in \Be-\Mg-\Be\, (Fig.~\ref{fig: MultimodeEITBeMgBe}(c)), the coupling drive is turned on at the same time as the EIT cooling beams, enabling cooling of the OOPH mode through the ALT mode in a continuously-coupled fashion. This is successful in part due to the broad bandwidth of EIT cooling; when the same procedure was attempted using Raman sideband cooling in our system, the cooling transition was dressed by the coupling drive which decreased the sideband cooling efficacy~\cite{Modecouplingforcooling}.

\section{Conclusion}
In this work, we demonstrate that EIT cooling can be adapted for systems with a more complex level structure than the prototypical $\Lambda$ level structure, arising here due to non-zero nuclear spin. This is especially important for applications where the coolant ion\textemdash either in a mixed-species ion crystal or as the only ion species being used\textemdash is required to perform additional functions where internal state coherence is beneficial; \Mg~is suitable for such a role. Using \Mg~as the coolant ion, all axial modes of both a \Be-\Mg~and a \Be-\Mg-\Be~ion crystal were cooled to $\bar{n}\leq 1$ in less than 1.3 ms, with most modes having $\bar{n}\leq 0.4$. It was also shown that while the presence of ion micromotion is detrimental to EIT cooling, the technique is still successful and is able to cool with reasonable efficiency, reaching occupations as low as $\bar{n} = 0.26\pm0.02$. We anticipate that much lower motional occupation can be achieved in experiments with lower micromotion, as suggested by numerical simulation.

During the preparation of this manuscript, we became aware of related work on EIT cooling of ions with nonzero nuclear spin, specifically $^{137}$Ba$^+$~\cite{Huang2024}. 

The authors thank C. M. Bowers and I. H. Zimmermann for helpful comments on the paper. J. J. W., P.-Y. H., S. D. E., Y. W., and G. Z. acknowledge support from the Professional Research Experience Program operated jointly by the National Institute of Standards and Technology (NIST) and the University of Colorado. S. D. E. acknowledges support from a National Science Foundation Graduate Research Fellowship (Grant No. DGE 1650115). A. D. B. and D. C. C. acknowledge support from National Research Council postdoctoral fellowships. This work was supported by IARPA and the NIST Quantum Information Program.

\bibliography{references.bib}
\vfill
\pagebreak
\section{Appendix A: Double (Dark) EIT Cooling}\label{sec: double dark EIT}
Laser-cooled systems with complex state structure such as \Mg~can also be used for extensions to EIT cooling, by making use of a ``double EIT" \cite{DoubleEIT} configuration, which should enable a lower final temperature for the mode of interest. This requires the addition of a third laser beam; instead of being used as a repump as in the main text, the third laser beam can be set at a similar detuning to the pump and probe laser in order to create an additional interference feature in the absorption spectrum. In contrast to the work shown in Ref.~\cite{DoubleBrightEIT} where a second mode is cooled by a second bright resonance, this uses the third beam to better cool a single mode.

In addition to the primary transparency feature that the $\Lambda$-EIT system creates, a second transparency feature can be created by blue-detuning this third beam to $\delta - \nu$, where $\delta$ is the detuning of the first two EIT beams and $\nu$ is the motional frequency. This should create a second absorption minimum at the motion-adding scatter frequency, which may suppress the dominant motion-increasing effect during regular EIT cooling. This should reduce the expected final temperature to $\mathcal{O}(\eta^2)$, where $\eta$ is the Lamb-Dicke parameter.

\begin{figure}[!h]
    \centerline{\includegraphics[width=0.4\textwidth]{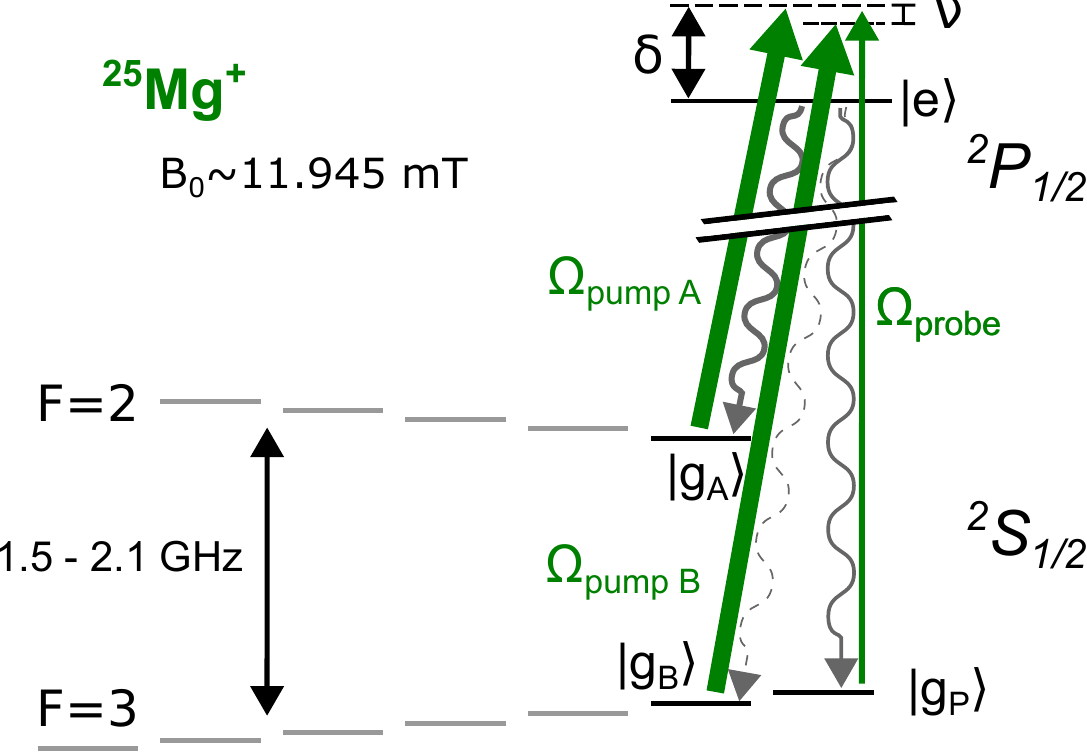}}
    \caption{Diagram of double EIT configuration for \Mg. Double EIT is performed by using two pump beams (labelled above as ``pump A" and ``pump B") with similar powers. One of the beams is detuned from resonance by $\delta - \nu$ (where $\nu$ is the frequency of the motional mode of interest) while the other two are detuned by $\delta$.}
    \label{fig: doubleEITdiagram}
\end{figure}

The third beam is typically assumed to have similar power to the original pump beam. Therefore, in this section we will refer to the original pump beam as ``pump A" and the repump beam as ``pump B". For double EIT, the condition that must be satisfied is~\cite{DoubleEIT}
\begin{equation}
\nu = (\Omega_{\rm pump A}^2 + \Omega_{\rm probe}^2 + \Omega_{\rm pump B}^2/2)/{4\delta},
\end{equation}
where $\Omega_{\rm pump A}$ and $\Omega_{\rm pump B}$ characterize the intensities of the pump A and pump B beams as described in Appendix B.

\begin{figure}[!h]
    \centerline{\includegraphics[width=0.48\textwidth]{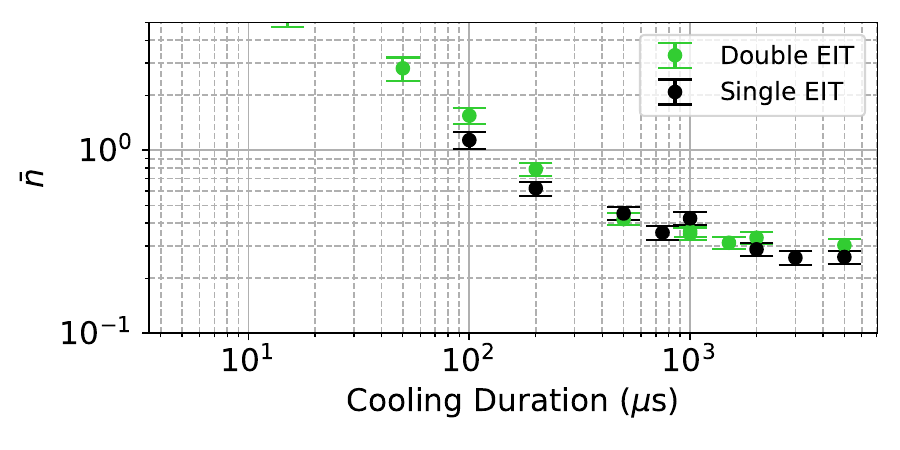}}
    \caption{Average motional mode occupation after double-EIT cooling (green points) compared to single-EIT-with-repump cooling as described in the main text (black points). }
    \label{fig: doubleEIT}
\end{figure}

Double-EIT cooling was attempted in our experiment on a single \Mg~ion crystal with axial frequency $\nu = 2\pi \times 1.18$ MHz. Unfortunately, the expected significant speedup \cite{DoubleEIT} compared to single-EIT-with-repump was not observed, as seen in Fig.~\ref{fig: doubleEIT}. We believe that this is due to the uncompensatable axial micromotion in our trap. For the double-EIT experiments, the probe and pump A beams were detuned to $+180$ MHz, while pump B was detuned to $+178.82$ MHz for optimal cooling of the axial mode at frequency $1.18$ MHz. It was also found experimentally through calibrations that a lower thermal occupation was reached when pump B was significantly weaker in intensity compared to pump A. The independently calibrated laser amplitudes were $\Omega_{\rm pump A}/2\pi = 26.33 \pm 0.19$ MHz, $\Omega_{\rm pump B}/2\pi = 1.12\pm0.10$ MHz, and $\Omega_{\rm probe}/2\pi = 4.11\pm0.32$ MHz. 

\section{Appendix B: Calibrating EIT Cooling}\label{sec: calibrations}\subsection{Optimizing EIT Cooling}
All three EIT beams were derived from the same laser (which was also used for resonant driving of the \textsuperscript{2}S\textsubscript{1/2} $\ket{F=3, m_F=2}$ $\rightarrow$ \textsuperscript{2}P\textsubscript{1/2} $\ket{3,3}$ transition), whose frequency was locked to a molecular iodine line. 

The power of each beam was individually controlled by an acousto-optic modulator (AOM), which also enabled the desired detunings to be set accurately. The rf power supplied to each AOM was varied in order to change the power of the beams. However, feedback stabilization of the beam power was difficult to implement due to space and beam power constraints, so the beam power would vary if the output power of the UV source drifted over time. The beams were fiber-coupled after the AOMs, so beam-pointing fluctuations at the fiber input also caused power fluctuations at the ion. Finally, some of the AOMs had temperature-dependent diffraction efficiency, which caused an additional variation in beam power when the duty cycle of the experiment was changed. Standard experimental duty-cycle management methods reduced but did not eliminate these power variations. These various forms of power fluctuations were suspected to be the main sources of error for the experiment, as changes in the beam power would change the absorption spectrum of EIT, thus changing the cooling rate.

For initial calibration of EIT cooling with a single \Mg~ion, $\Omega_{\rm pump}$ was first set to a value slightly less than $\sqrt{4\,\delta\,\nu}$ (recall that the EIT condition is given by $\nu = (\Omega_{\rm pump}^2 + \Omega_{\rm probe}^2)/{4\,\delta}$), using an EIT detuning of approximately $\delta = 2\pi \times 180$ MHz and a motional frequency of either $\nu = 2\pi \times 1.18$ MHz or $\nu = 2\pi \times 1.5$ MHz. This ensured that the majority of the dressing was done by the pump beam, and to minimize depumping towards the negative spin side of the \textsuperscript{2}S\textsubscript{1/2} manifold.

To account for the Stark shift from the micromotion sidebands of the pump beam on the common excited state \textsuperscript{2}P\textsubscript{1/2}$\ket{F=3, m_F=3}$, the detuning of the probe beam was initially adjusted above $\delta$ by a calculated value, typically 0.3-0.8 MHz. The power of the pump beam was first scanned and set at the value yielding optimal cooling using this calculated probe beam detuning. Next, the probe beam detuning was scanned to further optimize the cooling. This additional probe beam detuning did not require re-optimization over time. We found that it was necessary to optimize the pump beam power prior to scanning the probe beam frequency, as otherwise power drifts of this beam would make it difficult to reach a stable parameter set.

After these initial calibrations, the beam powers were optimized to yield the lowest motional mode occupation after 5 ms of EIT cooling with all beams on. First, the pump beam power was optimized with the probe and repump beams set to the expected powers from previous iterations ($\Omega_{\rm probe}/2\pi\approx 5$ MHz and $\Omega_{\rm repump}/2\pi\approx 1$ MHz). Next, the probe beam frequency was optimized; this step was only performed before a stable beam frequency was reached and was often not needed. The repump and probe beam powers were then optimized in turn; the final occupation was not found to be very sensitive to these parameters. 

For multi-ion cooling experiments, a similar procedure was performed, where each beam power was varied in turn. For the pulsed cooling method, the parameter set for each mode was found by optimizing for the lowest temperature of that mode, while all other axial modes were cooled using sideband cooling on \Mg\, (using interleaved mode-coupling swap operations to access the OOPH mode) to remove any extraneous effects in the measurement such as Debye-Waller factors~\cite{ExperimentalIssues}. For the simultaneous cooling method, the parameter set was chosen to obtain (approximately) the lowest total temperature of all modes.

The order of the modes to be cooled in the pulsed multi-ion cooling scans was chosen so that the mode with the highest heating rate was cooled last, to minimize the amount of time in which the mode could heat up again. The minimum duration required for each mode to reach its steady-state temperature was used, with the final pulse stopped before the steady state was reached to reduce the re-heating of the previously cooled modes (see main text, Fig.~\ref{fig: MultimodeEITBeMg}(b) and~\ref{fig: MultimodeEITBeMgBe}(b)). Iterating through the modes multiple times did not further decrease the final temperatures.

\subsection{Measurement of Experimental Parameters}
The micromotion modulation index was measured using Raman beams. Radial micromotion was minimized by adjusting voltages on the dc trap electrodes, leaving predominantly axial micromotion. The axial micromotion was then measured by driving the carrier ($n=0$), first-order ($n=1$), and second-order ($n=2$) micromotion sidebands of a Raman transition between hyperfine states. The relative Rabi frequencies of these transitions were then compared to determine the micromotion modulation index $\beta=1.17$ for the Raman beams. Because the Raman beams have a difference wavevector along the axial direction, while the EIT beams have a difference wavevector at a 45$^{\circ}$ angle to the axial direction, the micromotion modulation index of the EIT beams must be scaled by $1/\sqrt{2}$, giving $\beta=0.827$.

The beam parameters $\Omega_{\rm pump}$, $\Omega_{\rm probe}$, and $\Omega_{\rm repump}$ were determined by measuring the ac Stark shifts on the  \textsuperscript{2}S\textsubscript{1/2}$\ket{2,2}$, $\ket{3,2}$, and $\ket{3,3}$ states using Ramsey interferometry. The measured frequency shift $\Delta f_{\rm Stark}$ is related to the quoted $\Omega$ values as
\begin{equation}\label{eq: starkpowermeasurement}
 2\pi\Delta f_{\rm Stark} =  \sum_j\sum_{n=-5}^5 \left(\frac{J_{|n|}(\beta)}{J_{0}(\beta)}\right)^2\frac{\Omega^2}{4\delta_{n\,j}}\, , 
\end{equation}
where $\beta=0.827$ is the measured micromotion modulation index, $n$ is the order of micromotion sideband (calibrations assumed the total Stark shift was generated by each beam plus its micromotion sidebands up to $\pm 5^\mathrm{th}$ order; sidebands beyond this order should be negligible for $\beta=0.827$), $j$ indexes all the allowed transitions between the pair of states whose transition shift is being measured and states in the \textsuperscript{2}P\textsubscript{1/2} manifold, and $\delta_{n\,j}$ is the detuning of the $n$th micromotion sideband from the transition indexed by $j$. The factor of $J_{0}(\beta)$ normalizes the extracted $\Omega$ so that it is equivalent to an on-resonance Rabi frequency of the carrier only in our experiment (i.e. under micromotion). This choice of definition means that one can directly compare $\Omega$ values across cases either with or without micromotion.

The pump beam intensity was determined from the frequency shift on the $\textsuperscript{2}\mathrm{S}\textsubscript{1/2}\ket{3,2}\leftrightarrow\textsuperscript{2}\mathrm{S}\textsubscript{1/2}\ket{3,3}$ transition, which due to the $\sigma^+$ beam polarization only produces an ac Stark shift on the \textsuperscript{2}S\textsubscript{1/2}$\ket{3,2}$ state.
The probe beam intensity was measured from the shift on the \textsuperscript{2}S\textsubscript{1/2}$\ket{2,2}\leftrightarrow \textsuperscript{2}S\textsubscript{1/2}\ket{3,3}$ transition, which is primarily due to the ac Stark shift on \textsuperscript{2}S\textsubscript{1/2}$\ket{3,3}$ but contains contributions from the shift on \textsuperscript{2}S\textsubscript{1/2}$\ket{2,2}$. The +180 MHz detuning of the probe beam from resonance with the \textsuperscript{2}S\textsubscript{1/2}$\ket{3,3}\leftrightarrow\textsuperscript{2}P\textsubscript{1/2}\ket{3,3}$ transition, and its relatively low power, kept population loss during the Ramsey sequence manageable. Similar to the pump beam, the repump beam intensity was determined from the frequency shift on the $\textsuperscript{2}S\textsubscript{1/2}\ket{2,2}\leftrightarrow\textsuperscript{2}S\textsubscript{1/2}\ket{3,3}$ transition, in which only the shift on the $\textsuperscript{2}S\textsubscript{1/2}\ket{2,2}$ state was considered.

In some cases, the beam was either too intense or too weak to allow for Stark shift determination. In these cases, a depumping experiment was performed. First, the \Mg~state was prepared in the nominal ground state of the measured beam (for example, the \textsuperscript{2}S\textsubscript{1/2}$\ket{3,2}$ state for the repump). Next, the beam was turned on, and the rate at which the population left the prepared state was observed. This was then compared to simulation to find the beam intensity within $\approx10-20\% $ uncertainty.

For the simulations shown in Fig. \ref{fig: SingleMgEIT}(a), the tripod-with-micromotion simulation was performed with the measured experimental parameters, including the presence of micromotion sidebands on the EIT beams. The tripod simulation without micromotion used the pump, probe, and repump power in the $n=0$ carrier only (and did not include any micromotion sidebands). The $\Lambda$-EIT simulation used the measured probe carrier power with a calculated ideal pump power corresponding to this probe power, and did not consider the $\ket{g_B}$ state.

\subsection{Beam Powers for Multimode Cooling}

For simultaneous cooling in the \Be-\Mg\,crystal, the beam parameters were $\Omega_{\rm pump}/2\pi = 29.3 \pm 0.5$ MHz, $\Omega_{\rm probe}/2\pi = 4.25 \pm 0.03$ MHz, and $\Omega_{\rm repump}/2\pi = 1.51 \pm 0.03$ MHz. The beam parameters used for interleaved cooling of \Be-\Mg\,are shown in Table \ref{tab: BM interleaved powers}.
\begin{table}[t]
    \centering
   \begin{tabular}{c|c|c}
          \hline
          Mode & $\Omega_{\rm pump}/2\pi$ (MHz) & $\Omega_{\rm probe}/2\pi$ (MHz) \\
          \hline
          INPH & $31.7\pm 0.4$ & $3.80\pm 0.05$ \\
          \hline
          OOPH & $44.0\pm 0.4$ & $4.32\pm 0.04$ \\
          \hline
    \end{tabular}
    \caption{The beam parameters for interleaved cooling of the \Be-\Mg-\Be\, ion crystal, with $\Omega_{\rm repump}/2\pi = 1.47 \pm 0.03$ MHz for both cooling settings.}
    \label{tab: BM interleaved powers}
\end{table}

\begin{table}[t]
    \centering
   \begin{tabular}{c|c|c}
          \hline
          Mode & $\Omega_{\rm pump}/2\pi$ (MHz) & $\Omega_{\rm probe}/2\pi$ (MHz) \\
          \hline
          INPH & $32.2\pm 0.5$ & $3.78\pm 0.05$ \\
          \hline
          ALT & $50.13\pm 0.22$ & $5.39\pm 0.05$ \\
          \hline
    \end{tabular}
    \caption{The beam parameters for interleaved cooling of the \Be-\Mg-\Be\,ion crystal, with $\Omega_{\rm repump}/2\pi = 1.47 \pm 0.03$ MHz for both cooling settings.}
    \label{tab: BMB interleaved powers}
\end{table}

The beam parameters used for interleaved cooling of \Be-\Mg-\Be~are shown in Table \ref{tab: BMB interleaved powers}. For simultaneous cooling in the \Be-\Mg-\Be\,crystal, the beam parameters were $\Omega_{\rm pump}/2\pi = 40\pm 5$ MHz, $\Omega_{\rm probe}/2\pi = 4.64 \pm 0.06$ MHz, and $\Omega_{\rm repump}/2\pi = 1.64 \pm 0.03$ MHz.

\end{document}